\documentclass[
reprint,
superscriptaddress,
amsmath,amssymb,
aps,
pra,
longbibliography,
]{revtex4-1}
\usepackage{float}
\usepackage{xcolor}

\usepackage{graphicx}
\usepackage{dcolumn}
\usepackage{bm}
\usepackage[colorlinks=true, pdfstartview=FitV, linkcolor=blue, citecolor=blue, urlcolor=blue]{hyperref}
\usepackage{dsfont}

\usepackage[separate-uncertainty=true]{siunitx}
\usepackage[siunitx]{circuitikz}

\graphicspath{{figures/}}

\begin{document}

\preprint{APS/123-QED}

\title{A biased Ising model using two coupled Kerr parametric oscillators with external force}

\author{Pablo \'Alvarez}
\affiliation{Laboratory for Solid State Physics, ETH Z\"{u}rich, CH-8093 Z\"urich, Switzerland.}
\author{Davide Pittilini}
\affiliation{Laboratory for Solid State Physics, ETH Z\"{u}rich, CH-8093 Z\"urich, Switzerland.}
\author{Filippo Miserocchi}
\affiliation{Laboratory for Solid State Physics, ETH Z\"{u}rich, CH-8093 Z\"urich, Switzerland.}
\author{Sathyanarayanan Raamamurthy}
\affiliation{Laboratory for Solid State Physics, ETH Z\"{u}rich, CH-8093 Z\"urich, Switzerland.}
\author{Gabriel Margiani}
\affiliation{Laboratory for Solid State Physics, ETH Z\"{u}rich, CH-8093 Z\"urich, Switzerland.}
\author{Orjan Ameye}
\affiliation{Department of Physics, University of Konstanz, D-78457 Konstanz, Germany.}
\author{Javier del Pino}
\affiliation{Institute for Theoretical Physics, ETH Z\"{u}rich, CH-8093 Z\"urich, Switzerland.}
\author{Oded Zilberberg}
\affiliation{Department of Physics, University of Konstanz, D-78457 Konstanz, Germany.}
\author{Alexander Eichler}
\affiliation{Laboratory for Solid State Physics, ETH Z\"{u}rich, CH-8093 Z\"urich, Switzerland.}
\affiliation{Quantum Center, ETH Zurich, CH-8093 Zurich, Switzerland}

\date{\today}

\begin{abstract}
  Networks of coupled Kerr parametric oscillators (KPOs) are a leading physical platform for analog solving of complex optimization problems. These systems are colloquially known as ``Ising machines''. We experimentally and theoretically study such a network under the influence of an external force. The force breaks the collective phase-parity symmetry of the system and competes with the intrinsic coupling in ordering the network configuration, similar to how a magnetic field biases an interacting spin ensemble. Specifically, we demonstrate how the force can be used to control the system, and highlight the crucial role of the phase and symmetry of the force. Our work thereby provides a method to create Ising machines with arbitrary bias, extending even to exotic cases that are impossible to engineer in real spin systems.
\end{abstract}

	\maketitle



\begin{figure}[h!]
    \includegraphics[width=\columnwidth]{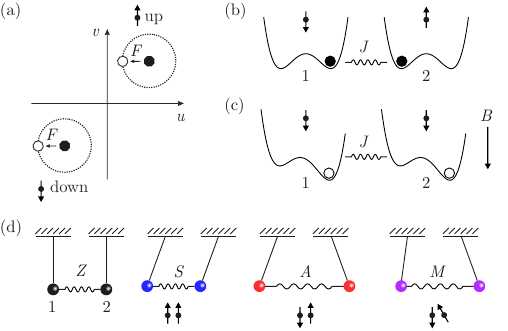}
    \caption{KPOs as Ising spin analogues. (a)~KPO phase states in a rotating phase space spanned by the quadratures $u$ and $v$, where $x=u\cos(2\pi f_d t) - v\sin(2\pi f_d t)$, with $f_d$ the driving frequency. Stationary solutions in the absence of an external force are shown as black dots and are labelled ``up'' and ``down'' to indicate a formal similarity to the two states of an Ising spin, marked by arrow symbols. Solution in the presence of an external force $F$ are shown as white dots as a function of the force phase $\theta$ (dotted circle). (b)~Representation of a 2-spin network as coupled double wells in the absence of a bias field. The levels of each double well ($1,2$) are energetically degenerate. The spin-spin coupling is indicated by $J$ and leads to an antisymmetric state in this example. (c)~In the presence of a bias field $B$, the degeneracy is broken and each spin has a preferred polarization. This preference can overcome the solution favored by the coupling $J$. (d)~Coupled resonators are often described in a normal-mode basis as illustrated by two coupled pendula. For two KPOs, the relevant nonlinear stationary states we will refer to in this paper include states with zero amplitude (Z), symmetric (S) and antisymmetric (A) oscillation, as well as mixed-symmetry (M) solutions.}
    \label{fig:fig1}
\end{figure}

The Kerr parametric oscillator (KPO) is a nonlinear resonator with a time-dependent harmonic potential term~\cite{Ryvkine_2006, Mahboob_2008, Wilson_2010, Eichler_2011_NL, eichler2018parametric, Gieseler_2012, Lin_2014, Puri_2017, Eichler_2018, Nosan_2019, Frimmer_2019, Grimm_2019, wang_2019, Puri_2019_PRX, Miller_2019_phase, yamaji_2022}. 
In a certain range of parameters, this potential modulation renders the zero-amplitude solution unstable. There, the system undergoes a spontaneous period-doubling $Z_2$ symmetry-breaking phase transition and assumes a large-amplitude oscillation at a so-called phase state. Importantly, a KPO features two such phase states with equal amplitude but phases separated by $\pi$, shown as black dots in Fig.~\ref{fig:fig1}(a).

The KPO is at the focus of much research work because its two phase states are analogous to the two polarization states of an Ising spin, ``up'' and ``down''. Consequently, it was proposed that networks of KPOs can be used to find the ground state of Ising Hamiltonians, that is, the energetically preferred configuration of a spin network~\cite{Ising_1925}. Such resonator-based Ising solvers~\cite{Mahboob_2016, Goto_2016, Puri_2019_PRX, Bello_2019, Okawachi_2020} are of high interest because the corresponding calculations are NP-hard to tackle with conventional computers~\cite{mohseni2022ising}, and yet they map to many other key optimization problems, such as the travelling salesman problem~\cite{Lucas_2014}, the MAX-CUT problem~\cite{Inagaki_2016_Science, Goto_2019}, and the number partitioning problem~\cite{Nigg_2017}. In Fig.~\ref{fig:fig1}(b), we sketch a network of two spins, where each spin takes the form of a double-well potential whose wells corresponds the two levels (``spin up'' or ``spin down'').

In the low-amplitude limit, the oscillations of a resonator network are approximately harmonic and can be understood in terms of their normal modes. A two-KPO system starting from the zero-amplitude (Z) solution can therefore ring up to symmetric (S) or antisymmetric (A) phase states that map to ferromagnetic or antiferromagnetic Ising configurations, respectively, see Fig.~\ref{fig:fig1}(d)~\cite{Heugel_2019_TC}. At large amplitudes, the nonlinearities become significant and can lead to deviations from a normal-mode basis. The deviations manifest for example as oscillations of mixed symmetry (M)~\cite{Heugel_2022}, which have no clear Ising counterpart. Nevertheless, the normal-mode basis remains useful as a frame of reference to study both strongly~\cite{Heugel_2022} and weakly coupled KPOs~\cite{Margiani_2023}.

\begin{figure*}[t]
    \includegraphics[width=\textwidth]{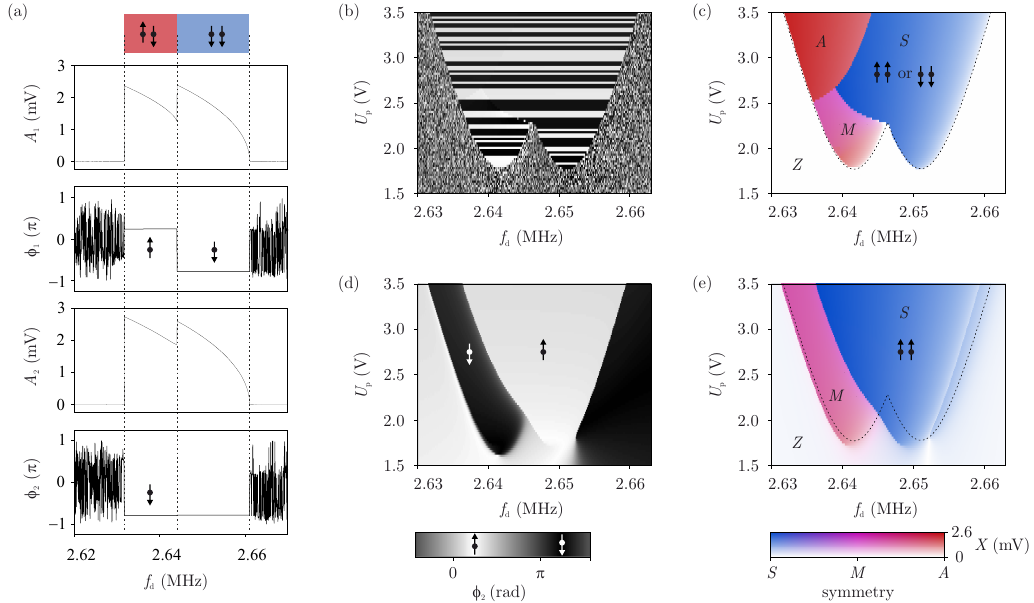}
    \caption{(a)~Measured amplitudes $A_i$ and phase responses $\phi_i$ of the two resonators when driven by a parametric pump tone with $U_p = \SI{3.5}{\volt}$. The arrow symbols represent the Ising spin analogy, and the colors indicate the relative phase configuration. (b)~Measured phase $\phi_2$ as a function of $f_d$ and $U_p$ for $F = 0$. (c)~Measured amplitude and phase configuration of the two resonators for $F = 0$. Dashed lines are theory predictions for the outline of the characteristic double-tongue. (d)~Same as (b) with a force strength $F = \SI{10}{\milli\volt}$ at a phase $\theta = \SI{55}{\degree}$. (e)~Measured amplitude and phase configuration of the two resonators for $F = \SI{10}{\milli\volt}$ and $\theta = \SI{55}{\degree}$. Dashed lines mark the outlines calculated for (c) to highlight the shifted boundaries.}
    \label{fig:fig2}
\end{figure*}

Previous experimental and theoretical studies of KPO-based Ising simulators with bilinear coupling considered the case of unbiased Ising Hamiltonians, where the solutions are only defined up to a global sign~\cite{Goto_2016,Puri_2017_NC,Nigg_2017,Goto_2018,Dykman_2018,Rota_2019,Strinati2019,Heugel_2022,Margiani_2023}. For example, the state ``down-up'' shown in Fig.~\ref{fig:fig1}(b) can be identically replaced by ``up-down'', as the individual spin levels are degenerate. Many Ising problems, however, require breaking of this degeneracy; the archetypal case being a magnetic field applied to a spin ensemble. In Fig.~\ref{fig:fig1}(c), a magnetic field $B$ biases the potential to compete with the spin-spin coupling $J$. A strong field can overcome the coupling-induced ordering and dramatically change the configuration that emerges as the optimal solution of the spin system. This functionality was recently included in optical parametric systems with dissipative coupling~\cite{Takesue_2020}, which constitute an alternative route to KPO networks~\cite{Inagaki_2016,gershenzon2020exact}.

In this paper, we demonstrate experimentally how the functionality of an external $B$ field can be introduced in a classical KPO network. By applying an external force $F$ to each resonator, we can displace the network's solutions in their phase space, as indicated in Fig.~\ref{fig:fig1}(a). Our experiment features two coupled KPOs, but the concepts we present are easily extended to larger networks. We provide a general framework to understand the role of the external force term for coupled KPOs and resonator-based Ising solvers. Importantly, while this forcing can emulate the effect of a simple bias field, we show that its consequences are much richer due to the freedom of selecting the driving amplitude and phase for each resonator individually. Our study will therefore not only provide a practical guide for applications, but also motivate further fundamental research on Ising networks with inhomogeneous magnetic fields, including quantum implementations~\cite{Puri_2017_NC,Grimm_2019,Dykman_2018,Goto_2016}.

Our experiment consists of two electrical resonators that feature a nonlinear capacitance~\cite{Nosan_2019}. The system can be described by the coupled differential equations
\begin{align}
	&\ddot{x}_i + \omega_0^2\left[1-\lambda\cos\left(4\pi f_d t\right)\right]x_i + \beta x_i^3 + \Gamma \dot{x}_i - Jx_i = F_i(t)\,,\label{eq:EOM}
\end{align}
where the displacement $x_{i}$ is the measured voltage signal of resonator $i$~\cite{Nosan_2019,Heugel_2022} as a function of time $t$. The angular resonance frequency $\omega_0/2\pi = f_0 = \SI{2.646}{\mega\hertz}$, the damping rate $\Gamma = \frac{\omega_0}{Q} = \SI{12}{\kilo\hertz}$, and the nonlinearity $\beta$ are approximately equal for both resonators, see section S1 in the supplemental material~\cite{Supplement}. Note that the capacitive diodes used to generate the nonlinearity $\beta$ mainly possess a quadratic force component (equivalent to the three-wave mixing enabled by $\chi^{(2)}$ in nonlinear optics), which allows us to implement parametric pumping $\lambda$ by driving the system with a voltage $U_p$ at the frequency $2f_d$~\cite{Nosan_2019}. The effective model we obtain in a frame rotating at $2\pi f_d$ is equivalent to that resulting from the well-known form in Eq.~\eqref{eq:EOM}~\cite{Eichler_Zilberberg_book}. Finally, $J$ quantifies the coupling between the resonators ($j\neq i$), and the $F_i(t)$ are external forces applied to the resonators individually. In the following, we will study the competition between these two effects ($J$ versus $F_i$) in ordering the phases of the two KPOs.

In Fig.~\ref{fig:fig2}(a), we demonstrate a frequency sweep by slowly increasing $f_d$ at a constant pump amplitude $U_p$ with $F_i = 0$. We use lock-in amplifiers (Zurich Instruments MFLI) to measure the amplitudes $A_i = (u_i^2 + v_i^2)^{1/2}$ and phases $\phi_i = \arctan(v_i/u_i)$ of the two resonators, where $x=u\cos(2\pi f_d t) - v\sin(2\pi f_d t)$. We observe that both resonators jump from zero to a finite amplitude around \SI{2.63}{\mega\hertz}. The phase of the two resonators assumes a well-defined value after the jump, with $\phi_1 - \phi_2 = \pi$. In our spin analogy, this corresponds to an antisymmetric ordering (``up-down''). Around \SI{2.645}{\mega\hertz}, the amplitudes jump again, accompanied by a $\pi$-shift of $\phi_1$. The resulting state is analogous to a symmetric spin state (``down-down''). Roughly at \SI{2.66}{\mega\hertz}, the amplitudes drop to zero and the phases become random again.

We repeat frequency sweeps at different values of $U_p$. In Fig.~\ref{fig:fig2}(b), the measured $\phi_2$ as a function of $f_d$ and $U_d$ is plotted. We clearly see that in each sweep, the phase assumes a well-defined value in a certain frequency range described by two overlapping, rounded triangles often called ``Arnold tongues''~\cite{Heugel_2022,Eichler_Zilberberg_book}. Importantly, the phase randomly assumes one of two values separated by $\pi$ in each sweep. This randomness is a consequence of the symmetry illustrated in Fig.~\ref{fig:fig1}(b), and the resulting spontaneous time-translation symmetry breaking at each jump from zero to finite amplitude. It is a fundamental feature of a KPO, and of networks thereof, in the absence of an external force.

To map the different symmetry phases of the system, we employ the symmetric and antisymmetric quadratures $v_S = \frac{1}{\sqrt{2}}(v_1 + v_2)$ and $v_A = \frac{1}{\sqrt{2}}(v_1 - v_2)$. Note that $u_{S,A}$ can be defined analogously and yield qualitatively similar results. In Fig.~\ref{fig:fig2}(c), we represent the total amplitude $X = (v_S^2 + v_A^2)^{1/2}$ as a brightness contrast, while the relative phase of the two resonators is shown in color code. As in a previous work~\cite{Heugel_2022}, we observe $S$, $A$ and $M$ phases inside the overlapping Arnold tongues. The center frequencies and phase symmetries of the two Arnold tongues are inherited from the normal modes of the system in the linear regime. The precise shapes of these zones are understood to originate from an interplay of the nonlinearity $\beta$ and the coupling $J$, and can be reproduced precisely with numerical simulations and with an analytical solver~\cite{kovsata2022harmonicbalance,Supplement}.

In a next step, we additionally apply an external force to each resonator. For simplicity, we select $F_i(t) = F \cos(2\pi f_d t + \theta)$ for both resonators, with $\theta$ a global phase. In Fig.~\ref{fig:fig2}(d), we see a striking change in the measured $\phi_2$ compared to Fig.~\ref{fig:fig2}(b): the phase now exhibits a well-defined value over the entire parameter space, and the jump between solutions follows a deterministic pattern. The spontaneous time-translation symmetry breaking observed in Fig.~\ref{fig:fig2}(b) is entirely replaced by a force-induced bias, as sketched in Fig.~\ref{fig:fig1}(c). The force therefore allows us to control the ``spin polarization'' of the individual KPOs in each sweep.

\begin{figure}[t]
    \includegraphics[width=\columnwidth]{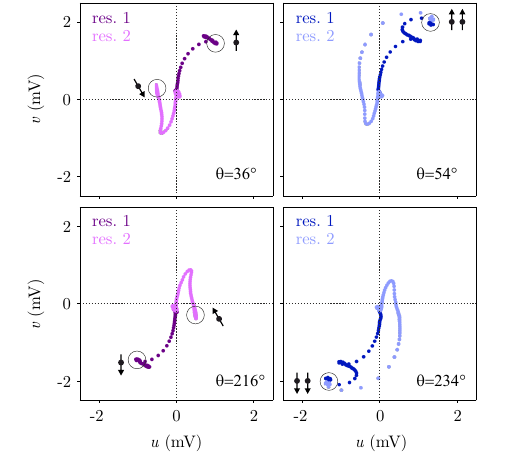}
    \caption{Ringup experiments with different values of force phase $\theta$. Both resonators are initialized in $u_i=0$ and $v_i=0$ (center of graphs). First, only an external force $F=\SI{20}{\milli\volt}$ with phase $\theta$ is applied to break the symmetry, bringing the resonators to an amplitude of $\approx \SI{0.1}{\milli\volt}$. Then the parametric pump $U_p=\SI{3.15}{\volt}$ is added and the resonators ring up to symmetric or mixed-symmetry solutions, as marked by white disks. Arrow symbols indicate the corresponding spin analogy. All experiments were performed at $f_d = \SI{2.639}{\mega\hertz}$.}
    \label{fig:fig3}
\end{figure}

\begin{figure*}
    \includegraphics[width=\textwidth]{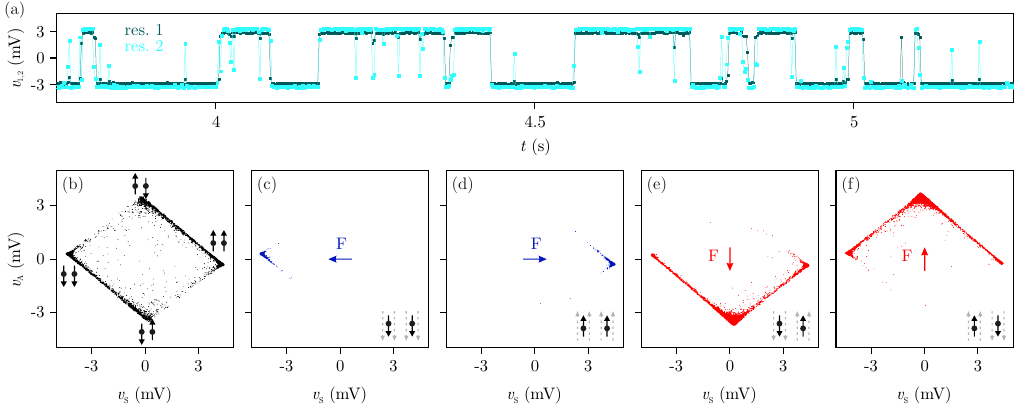}
    \caption{Stochastic sampling of KPO network states. (a)~Quadratures $v_1$ and $v_2$ measured as a function of time in the presence of force noise. The applied noise is white up to approximately \SI{30}{\mega\hertz} and has a standard deviation of \SI{350}{\milli\volt}. Separate noise generators were employed for the two KPOs and connected via additional inductive lines. Only the $v_i$ are shown here for simplicity. $U_p = \SI{3.5}{\volt}$, $F = 0$, lock-in demodulation rate $B = \SI{500}{\hertz}$, and $f_d = \SI{2.586}{\hertz}$ (the resonance frequencies of both KPO shifted due to the additional inductive lines). (b)~Representation of the timetrace from (a) in a so-called symmetry space spanned by $v_S = \frac{1}{\sqrt{2}}(v_1 + v_2)$ and $v_A = \frac{1}{\sqrt{2}}(v_1 - v_2)$. (c)~A symmetric force of \SI{30}{\milli\volt} with $\theta = 0$ (after subtracting a global phase offset) favors the negative symmetric state. (d)~With $\theta = \SI{180}{\degree}$, the symmetric force favors the positive symmetric state. (e)~An antisymmetric force with $\theta_1 = 0$ and $\theta_2 = \SI{180}{\degree}$ favors the negative antisymmetric state. (f)~inverting both phases favors the positive antisymmetric state. A total of \SI{60}{\second} of data was taken for each case. Insets show spin with grey dashed arrows for the local magnetic field direction.}
    \label{fig:fig4}
\end{figure*}

When plotting the system states in Fig.~\ref{fig:fig2}(e), we find several important differences to the unbiased example in Fig.~\ref{fig:fig2}(c). First, the outlines of the Arnold tongues are shifted by the force. This effect arises due to changes in the stability conditions of the parametric oscillators, which was previously studied for the case of a single KPO~\cite{Ryvkine_2006,Rhoads_2010,Papariello_2016,Leuch_2016,eichler2018parametric,Nosan_2019}. The most prominent manifestation of this effect is a jump (of both phase and amplitude) at the right border of the $S$ phase. This jump is caused by the termination of the selected symmetric phase state~\cite{Supplement}. For a symmetric force, we mainly observe a shift of the $S$ region. Second, the antisymmetric state labelled $A$ is no longer visible, as it is not favored by the symmetric force. Numerical simulations indicate that the $A$ state still appears at higher values of $U_p$, where the impact of the external force is reduced relative  to the parametric pump. The symmetric state $S$ therefore fills a greater portion of the diagram, in agreement with the intuition that a symmetric force $F_1 = F_2$ should favor this phase configuration. Third, the $S$ state now only comprises an ``up-up'' component, instead of allowing both ``up-up'' and ``down-down'' as in Fig.~\ref{fig:fig2}(c). The force therefore fulfills the role of a potential bias, as anticipated.

In Fig.~\ref{fig:fig2}, we employ frequency sweeps with a fixed force phase $\theta$ to study the solution space of our system. In many applications, however, it may be more useful to start from the equilibrium state ($u_i = 0$ and $v_i = 0$), and then directly access a particular solution through the correct choice of $\theta$. We demonstrate this capability experimentally in Fig.~\ref{fig:fig3}, where the system rings up to four different symmetric and mixed-symmetry states, depending on the phase $\theta$ of the symmetric external force. In the spirit of the two-spin analogy that we employ throughout the paper, we label the final states with corresponding symbols. We find that small changes in $\theta$ can cause the system to attain an entirely different state. For instance, $\theta = \SI{36}{\degree}$ leads to a symmetric state while $\theta = \SI{54}{\degree}$ leads to a mixed-symmetry state. Rotating $\theta$ by \SI{180}{\degree} inverts the final state of both resonators, leaving the relative configuration intact (e.g. ``up-up'' becomes ``down-down''). 

The deterministic experiments we reported so far always reach the same stable solutions for the same initial conditions and driving parameters. This leads to a limited understanding of our system, which possesses several stable solutions for certain positions in the $f_d$-$U_p$ diagram~\cite{Supplement}. Stochastic sampling allows us to explore these different solutions, and to assess their dwell times in the long-time limit~\cite{Margiani_2023}. In Fig.~\ref{fig:fig4}(a), we show how the system jumps between different solutions when activated by white noise and in the absence of external forcing ($F_i = 0$). For most of the time, the system jumps between the two symmetric solutions, which are more stable than the antisymmetric solutions due to their higher amplitude. A high amplitude imposes a larger ``momentum barrier'' and makes the state more stable against jumps~\cite{Margiani_2023}. In a so-called symmetry space spanned by $v_S$ and $v_A$, this dynamics result in the plot shown in Fig.~\ref{fig:fig4}(b). Here, symmetric and antisymmetric states appear at the corners of a diamond at $v_A = 0$ and $v_S = 0$, respectively (plots for the $u_S$ and $u_A$ look very similar). The edges of the diamond result from points measured during transitions between the states. Below the graph, we sketch the corresponding picture of two spins without an external field, allowing both symmetric and antisymmetric states (and fluctuations between them).

When applying an external force, we can change the relative weight (occupation probability) of these four states. In Fig.~\ref{fig:fig4}(c) and (d), we apply symmetric forces whose phase $\theta$ is tuned to favor either of the two symmetric states. As a consequence, we observe the system only in the corresponding symmetric state, and all jumps are suppressed. The spin picture with external field below each graph shows how the spins are forced to align by the homogeneous external field.

In contrast to a real (nanoscale) spin ensemble, our system allows us to arbitrarily tune the amplitude and phase of the applied force (corresponding to the strength and direction of the external field) for each KPO individually. As a demonstration example, we show in Fig.~\ref{fig:fig4}(e) and (f) the results for an antisymmetric force $F_i(t) = F \cos(2\pi f_d t + \theta_i)$ with $\theta_2 = -\theta_1$. Here, we find that the system occupies almost exclusively the selected antisymmetric state. However, the suppression of the symmetric states is weaker than in the opposite case in Fig.~\ref{fig:fig4}(c) and (d). Again, this is due to the fact that the symmetric state is generally more stable than the antisymmetric state for $J/\beta >0$~\cite{Margiani_2023}. Even larger forces would be necessary to entirely overcome this intrinsic bias.

In our experiments and in the theory analysis, we find that the external force can bias our system of coupled KPOs. In the simplest case, this bias is analogous to an external magnetic field acting on an ensemble of coupled spins.
The tunable phase $\theta$ of the external force assumes the role of the magnetic field angle, which can be different for each KPO, cf. Fig.~\ref{fig:fig3} and Fig.~\ref{fig:fig4}. This freedom in selecting the phases will be crucial in future experiments that go beyond conventional spin systems, as it allows access to unexplored, exotic networks that have no counterpart in solid state physics. Such novel networks include, for instance, the Ising chain in the presence of a tunable local impurity~\cite{Falk_1966}, the random-field Ising model~\cite{Grinstein_1976,belanger1991random,Bingham_2021,yao2023thermal} and corresponding avalanche models~\cite{Percovic_1995,im2009direct,Field_1995,Lahini_2017}, and the magnetic Bose polaron \cite{mistakidis2022inducing}. We believe that controlled experimental realizations of these elusive phenomena will spur new developments in theory in many directions. At the same time, our work demonstrates new strategies to control Ising machines, and to program such systems to solve complex optimization tasks~\cite{mohseni2022ising,Lucas_2014,Inagaki_2016_Science, Goto_2019,Nigg_2017}.

\appendix

\section{Setup Calibration}

Before starting any experiments with the two coupled resonators, we calibrate their characteristics individually. The resonance frequencies of our circuits can be tuned via the bias voltages applied to the diodes across \SI{47}{\mega\ohm} resistors that act as a low-pass filters~\cite{Nosan_2019}. We can apply a voltage to one resonator while keeping the other resonator at zero bias, which results in a large detuning between the resonance frequencies. Driving the biased resonator with a small external drive enables us to extract its resonance frequency, while a sweep with a large drive allows us to assess its nonlinear coefficient. This procedure is then repeated with the second resonator after tuning the bias voltage until the resonance frequencies match within the experimental uncertainty, which is on the order of \SI{100}{\hertz}. We noted, however, that the device characteristics can change slightly when they are resonantly coupled. For this reason, we extract the precise parameters of the coupled system from the measured data, as described below.

\section{Numerical Simulations}


The boundary of the large-amplitude region in Fig.~2(c) corresponds to two overlapping Arnold tongues~\cite{Heugel_2022,Eichler_Zilberberg_book}. As an initial guess, we assume the two resonators to be identical. By fitting a line to this boundary, we can extract the parametric threshold $U_\mathrm{th} = \SI{1.78}{\volt}$ (the lowest tip of the tongues) and the quality factor $Q = \num{227}$, as well as a resonance frequency $f_0 = \SI{2.646}{\mega\hertz}$ (the middle between the two tongues) and frequency splitting $\Delta f = \SI{9.4}{\kilo\hertz}$ (the separation between the tongues). The coupling constant $J$ can then be calculated as $J = (2\pi)^2 f_0 \Delta f$, while the threshold $U_\mathrm{th}$ and the quality factor $Q$ allow us to transform the parametric drive amplitudes $U_d$ to unit-less parametric modulation depths $\lambda = \frac{2 U_d}{Q U_\mathrm{th}}$.

\begin{figure*}[t]
    \includegraphics[width=\textwidth]{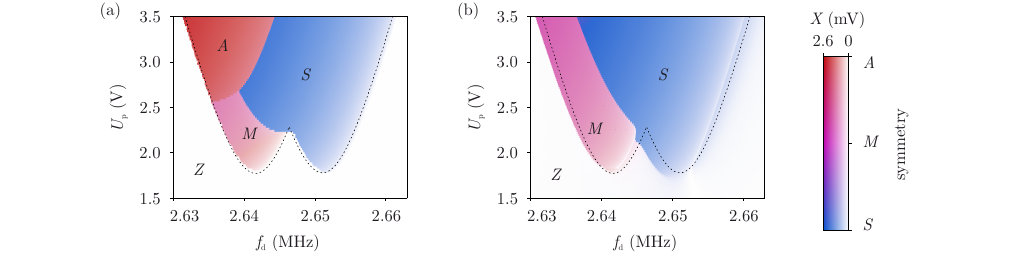}
    \caption{Amplitude and phase configuration (symmetry) of the two resonators calculated with a numerical simulation tool~\cite{Eichler_Zilberberg_book}. The results were matched to the measurements shown in Fig.~2 of the main text by fine-tuning the parameters. We obtained $\omega_1/2\pi = \SI{2.647}{\mega\hertz}$, $\omega_2/2\pi = \SI{2.645}{\mega\hertz}$, $J = \SI{0.98}{\mega\hertz\squared}$, $Q_1 = 234$, $Q_2 = 219$, $\beta_1 = \SI{-8.4e17}{\per\volt\squared\per\hertz\squared}$, and $\beta_2 = \SI{-6.8e17}{\per\volt\squared\per\hertz\squared}$. (a)~Simulation result for $F = 0$. The black line is identical with the one in Fig.~2(c). (b)~We reproduce the result in Fig.~2(e) with an effective force $F = \SI{1.6e+08}{\volt\per\square\hertz}$ and a phase $\phi = \SI{-40}{\degree}$. As for the experiment, the theoretically expected Arnold tongue boundaries (dashed lines) match in (a) but not in (b) due to the shifted boundaries.}
    \label{fig:figS1}
\end{figure*}

In a second step, we use these parameters to perform numerical simulations of frequency sweeps for different modulation depths $\lambda$. We find that the phase transitions between antisymmetric, symmetric and mixed dynamical phases  ($A$, $S$ and $M$, respectively) are highly sensitive to non-degeneracies bewteen the resonator parameters. We therefore recursively modify the individual resonance frequencies, quality factors, and nonlinear coefficients slightly to optimize the agreement with the measured results. This procedure yielded the parameters used for Fig.~\ref{fig:figS1}(a).

Finally, an external force term is added to the simulations. The applied force strength and phase depends on the induction parameters between the driving coils and the resonators coils, while the measurement gain and phase depend on the induction parameters between the resonator coils and the pickup coils. As these two sets of parameters could not be calibrated independently, we decided to determine the effective strength and phase of the drive $F$ by fine-tuning until the observed features are comparable to the measured results. This leads us to the values used in Fig.~\ref{fig:figS1}(b).

\section{Number and stability of Solutions}

For a numerical analysis of the KPO network, we employ the open-source software package HarmonicBalance.jl~\cite{kovsata2022harmonicbalance}. The package allows us to transform the coupled equations from Eq.~(1) into slow-flow equations for the quadratures $u_i$ and $v_i$:
\begin{align}
    \label{eq:slow_flow_2_coupled}
    \dot{u}_{1,2}^{\phantom\dagger} &= -\frac{\Gamma  u_{1,2}^{\phantom\dagger}}{2}- \left(\frac{3 \beta}{8 \omega }X_{1,2}^2 +\frac{\omega_0^2 - \omega^2}{2\omega}+\frac{\lambda \omega_0^2}{4 \omega }\right) v_{1,2}^{\phantom\dagger} \nonumber\\&+\frac{J  v_{2,1}^{\phantom\dagger}}{2 \omega} + \frac{F_{u,1,2}}{m}\,,\nonumber\\
    \dot{v}_{1,2}^{\phantom\dagger} &= -\frac{\Gamma  v_{1,2}^{\phantom\dagger}}{2} + \left(\frac{3 \beta}{8 \omega }X_{1,2}^2 + \frac{\omega_0^2 - \omega^2 }{2\omega} -\frac{\lambda \omega_0^2}{4 \omega }\right)u_{1,2}^{\phantom\dagger}\nonumber\\&-\frac{J  u_{2,1}^{\phantom\dagger}}{2 \omega} + \frac{F_{v,1,2}}{m}\,,
\end{align}
with $X_{i} = (u_i^2+v_i^2)^{1/2}$ the total amplitude of resonator $i$ and $F_{u,i}$, and $F_{v,i}$ the quadrature components of the force acting on it, respectively. Our main focus lies in examining the system's steady states, specifically the solutions of $u_{1,2}$ and $v_{1,2}$ when $\dot{u}_{1,2}=\dot{v}_{1,2}=0$. Therefore, the task of identifying all the steady states reduces to determining all the roots of the polynomial system described by Eq.~\eqref{eq:slow_flow_2_coupled}. To accomplish this, HarmonicBalance.jl utilizes Homotopy Continuation~\cite{HomotopyContinuation.jl}, a technique that enables the computation of these roots.

\begin{figure*}[t]
    \includegraphics[width=\textwidth]{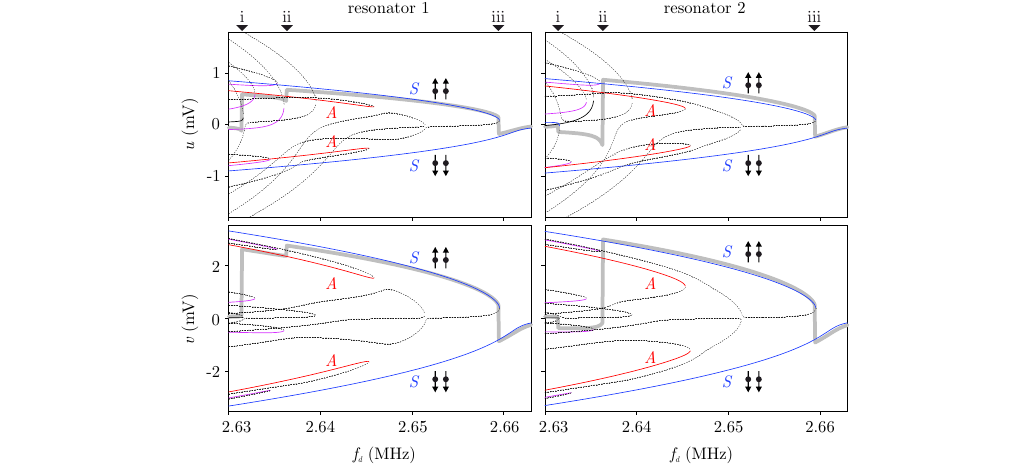}
    \caption{Comparison of measured quadratures $u$ and $v$ for resonator 1 and 2 as a function of frequency. Experimental data are shown as thick grey lines and are extracted from the measurement shown in Fig.~2(e) for $U_p = \SI{3.5}{\volt}$. Stable theory solutions are shown in blue for symmetric ($S$), red for antisymmetric (A), pink for mixed-symmetry (M), and black for the solution near zero amplitude ($Z$). Dashed lines represent unstable solutions. Arrow symbols indicate analogy to a two-spin system for thy symmetric solutions. The roman numbers i to iv indicate frequencies where the experimental trace jumps between solutions.}
    \label{fig:figS3}
\end{figure*}

We can evaluate all possible roots of Eq.~\eqref{eq:slow_flow_2_coupled} as a function of $\omega$ and $\lambda$, and assess their stability with standard methods~\cite{Eichler_Zilberberg_book}.
Comparing these solutions as a function of $f_d$ to an experimental sweep provides additional insights. An example comparison in the case with external force is plotted in Fig.~\ref{fig:figS3}. There, we can follow the measured system state (thick grey line) from low to high $f_d$, starting in a state near zero amplitude. Roughly at \SI{2.631}{\mega\hertz}, the system jumps to one out of four solutions with mixed symmetry (see arrow labelled $\textbf{i}$). These solutions are all different for non-identical resonators and respond sensitively to any tuning of the parameters, making it difficult to obtain a good agreement between theory and experiment without heuristic fine-tuning (which we avoided here).

At \SI{2.636}{\mega\hertz}, the system jumps from the mixed-symmetry solution to a symmetric solution ($\textbf{ii}$). From the theory analysis, we understand that a jump is necessary because the mixed-symmetry solutions all terminate in bifurcations where they merge with unstable solutions. In principle, the system offers two symmetric ($S$) and two antisymmetric ($A$) stable solutions that the system could jump to. We interpret the selected symmetric state as dictated by the external force, as already discussed in Fig.~2. Future analysis will be dedicated to analyzing the precise rules governing the state selection.

At \SI{2.66}{\mega\hertz}, the system jumps a third time, with both resonators undergoing a phase change of $\pi$ ($\textbf{iii}$). This situation is reminiscent to what was found in a single KPO in the presence of a symmetry-breaking force~\cite{Rhoads_2010,Papariello_2016,Leuch_2016,eichler2018parametric,Nosan_2019}. There, the external force resulted in a local amplitude minimum associated to a bifurcation, and to a jump in a frequency sweep. In the present case, we observe the same phenomenon for the symmetric normal mode of the system. The third jump at $\textbf{iii}$ is therefore a direct manifestation of the interplay between the parametric pump and the external force applied to a parametric oscillator network.

\providecommand{\noopsort}[1]{}\providecommand{\singleletter}[1]{#1}%

\end{document}